\documentstyle[12pt,epsfig]{article}
\begin{document}
\title{Quantum--Mechanical Detection of Non--Newtonian Gravity.}
\author{ A. Camacho
\thanks{email: acamacho@aip.de} \\
Astrophysikalisches Institut Potsdam. \\
An der Sternwarte 16, D--14482 Potsdam, Germany.}

\date{}
\maketitle

\begin{abstract}
In this work the possibility of detecting the presence of a Yukawa term,
as an additional contribution to the usual Newtonian gravitational potential,
is introduced.
The central idea is to analyze the effects at quantum level employing interference patterns
(at this respect the present proposal resembles the Colella, Overhauser and Werner experiment),
and deduce from it the possible effects that this Yukawa term could have.
We will prove that the corresponding inter\-fe\-rence pattern depends on the phenomenological parameters that define this kind of terms.
Afterwards, using the so called res\-tric\-ted path integral formalism, the case of a particle whose position is being continuously
monitored, is analyzed, and the effects that this Yukawa potential could have on the measurement outputs are obtained.
This allows us to obtain another scheme
that could lead to the detection of these terms. This last part also renders new
theo\-re\-tical predictions that could enable us to confront the res\-tric\-ted
path integral formalism against some future experiments.
\end{abstract}
\bigskip
\newpage
\section{Introduction}
\bigskip

General Relativity (GR) is one of the milestones of modern physics, and currently
many of its predictions have been already tested. For instance, we
already have the following results: gravitational time dilation measurement [1], gravitational deflection
of electromagnetic waves [2], time delay of electromagnetic waves in the field of the sun
[3], or the geodetic effect [4]. Neverwithstanding, at this point it
is also important to comment that all these impressive direct 
confirmations of GR are confirmations of weak field corrections to the Galilei--Newton mechanics.

The discovery of the first binary pulsar PSR1913+16 [5] allowed to probe the propagation
properties of the gravitational field [6], the results between theory and experiment agree at a level of $10^{-3}$. The possibilities that binary pulsars 
offer do not finish here, they can also be used as laboratories for testing strong--field gravity [7].
Concerning binary pulsars at this point it is noteworthy to mention that they are
a confirmation of general relativity done at the classical level, here we mean that
the observations and predictions comprise the orbital dynamics of a binary pulsar,
for instance, orbital period, eccentricity [8].

Therefore, if GR is so successful, then why should we need analyze some possible
deviation of the Newtonian inverse--square force law?. The answer stems from the
fact that the agreement between general relativity and experiment might be compatible
with the existence of a scalar contribution to gravity, such as a dilaton field [9].

This dilaton field emerges in several theoretical attempts that try to formulate
a unified theory of elementary particle physics. As one of their consequences they
predict the existence of new forces (which are usually refered to as ``fifth force''), whose effects extend over macroscopic distances [10].
In some ways, these new forces simulate the effects of gravity, but a crucial
point is that they are not described by an inverse--square law, and even
more, they, generally, violate the Weak Equivalence Principle (WEP) [11].
Hence the presence of this kind of forces, coexisting with gravity, could be
detected, in principle, by apparent deviations from the inverse--square law,
or from the violation of WEP. Hence, a strong theoretical motivation for analyzing
possible deviations from Newtonian gravity is to probe for new fundamental forces
in nature.

To date, after more than a decade of experiments [12], there is no compelling e\-vi\-den\-ce
for any kind of deviations from the predictions of Newtonian gravity.
But Gibbons and Whiting (GW) phenomenological analysis of gravity data [13] has
proved that the very precise agreement between the predictions of Newtonian gravity
and observation for planetary motion does not preclude the existence of large
non--Newtonian effects over smaller distance scales, i.e., precise experiments
over one scale do not ne\-ce\-ssarily constrain gravity over another scale.

GW results conclude that the current ex\-pe\-ri\-men\-tal constraints over possible
deviations did not severly test Newtonian gravity over the $10$--$1000$m distance
scale, usually called ``geophysical window''.

Recently, a new test of the equivalence principle was carried out [14], the one sets new constraints on the
possible ranges of a Yukawa term. This new experiment improves the current limit
for ranges between 10km and 1000km. Neverwithstanding, in the short range it can say
nothing about distances smaller than 1cm. Nevertheless, this experiment is performed
on a classical system, namely, a 3 ton $^{238}U$ attractor rotates around a torsion balance,
which contains Cu and Pb macroscopical test bodies. In this experiment the differential acceleration
of the test bodies toward the attractor was measured.

Finally, we must also mention the experiment been already carried out at
the University of Padova [15]. This proposal measures the displacements induced by an oscillating
mass acting as a source of gravitational field on a micromechanical resonator. This device
could give information about scalar interactions in the range below 1mm.

Among the models that in the direction of noninverse--square forces currently exist we have Fujii's proposal [16], in which a ``fifth force'', coexisting simultaneously with gravity, comprises a modified Newtonian
potential with a Yukawa term, $V(r) = -G_{\infty}{mM\over r}\left(1 + \alpha e^{-{r\over \lambda}}\right)$,
here $G_{\infty}$ describes the interaction between $m$ and $M$ in the
limit case $r\rightarrow\infty$, i.e., $G = G_{\infty}(1 + \alpha)$,
where $G$ is the Newtonian gravitational constant. This kind of deviation terms arise from 
the exchange of a single new quantum of mass $m_5$, the Compton wavelength
of the exchanged field is $\lambda = {\hbar\over m_5c}$ [17], this field is usually called dilaton.

If we take a look at the experimental efforts that have been done in order to
test the inverse--square law we will find that they can be separated into two
large classes: (i) those experiments which involve the direct measurement
of the magnitude $G(r)$, they compare preexisting laboratory Cavendish measurements
of $G$ [18]; and (ii) the direct measurement of $G(r)$ with $r$ [19]. A relevant
characteristic of these efforts has to be mentioned, they remain always at the
classical level, the action of the Yukawa term is always on classical systems,
namely, classical test masses (Cavendish case), or in the case of mine and
Borehole experiments, once again, classical test particles are employed.
One of the exceptions around this topic is the use of the Casimir effect [20],
here Planck constant, $\hbar$, appears as a parameter in the experiment, another quantum analysis
may be found in [21].
Neverwithstanding, the existence of retardation forces, such as van der Waals forces, complicates these classical experimental constructions [17].

In this work we will try to explore the theoretical predictions, at quantum level,
that a Yukawa term could have. This will be done resorting to an experimental
proposal which is very similar to the Colella, Overhauser, and Werner (COW) cons\-truc\-tion [22].
As we already know, COW allows the detection of the gravity--induced phase difference
between the amplitudes of two wave packets arriving at a certain detection point.
Having this in hindsight, we could wonder if the parameters $\alpha$ and $\lambda$,
appearing in Fujii's model, could render a detectable effect in this type of
COW experiment. In other words, we will calculate the non--Newtonian gravity--induced
interference between the amplitudes of two wave packets and compare this
result with COW. It will be found that the difference depends on $\alpha$ and $\lambda$,
and therefore could be detected, at least in principle. It gives also the possibility
of measuring the mass of the dilaton field. Afterwards, we will consider the continuous monitoring of the position of the
two particle beams and see that, in the context of the restricted path integral
formalism (RPIF) [23], $\alpha$ and $\lambda$ do appear explicitly in the
resulting interference term, and hence we obtain, comparing with the corresponding
results of the Newtonian case [24], an additional method to determine these non--Newtonian 
parameters.
\bigskip

\section{Non--Newtonian Gravity--Induced Interference}
\bigskip

As was already mentioned above, let us now consider the case of a Yukawa modification 
to the Newtonian gravitational potential [16]

{\setlength\arraycolsep{2pt}\begin{eqnarray}
V(r) = -G_{\infty}{mM\over r}\left(1 + \alpha e^{-{r\over \lambda}}\right).
\end{eqnarray}

The Lagrangian of a particle with mass $m$, moving in this field, is

{\setlength\arraycolsep{2pt}\begin{eqnarray}
L = {m\over 2}\dot{\vec{r}}^2 + G_{\infty}{mM\over r}\left(1 + \alpha e^{-{r\over \lambda}}\right).
\end{eqnarray}

Let us now write $r = R + l$, where $R$ is the Earth's radius, and $l$ the height
over the Earth's surface. Therefore, keeping terms up to second order in $l$, we find

{\setlength\arraycolsep{2pt}\begin{eqnarray}
L = {m\over 2}\dot{\vec{r}}^2 + G_{\infty}{mM\over R}
{\Bigg(}\left[1 + \alpha + {\alpha R\over\lambda}\left({R\over 2\lambda} - 1\right)\right] -
\left[{1 + \alpha\over R} - {\alpha R\over 2\lambda^2}\right]l + {1 + \alpha\over R^2}l^2{\Bigg)}.
\end{eqnarray}

Let us now consider the case in which we perform an
experiment similar to COW [22], i.e., two particles, starting
at point $P$, move along two different trajectories, $C$ and $\tilde{C}$, and afterwards
they are detected at a certain point $Q$. Here we assume that the size of the
wavelengths of the packets is much smaller than the size in which the field changes considerably (i.e., we
are always in the short wavelength limit), and in consequence we may consider a semiclassical approach in the analysis of the wave function.

Hence the wave function is given by the following expression

{\setlength\arraycolsep{2pt}\begin{eqnarray}
\psi(\vec{r}, t) \sim {1\over [E - V(\vec{r})]^{{1\over 4}}}
\exp\left\{{\pm{i\over\hbar}}\int_{(P)}^{(Q)}
\sqrt{2m[E - V(\vec{r})]}d\tilde{L} - {i\over\hbar}Et\right\},
\end{eqnarray}

\noindent where $V(\vec{r}) = -G_{\infty}{mM\over R}
{\Bigg(}[1 + \alpha + {\alpha R\over\lambda}({R\over 2\lambda} - 1)] -
[{1 + \alpha\over R} - {\alpha R\over 2\lambda^2}]l + {1 + \alpha\over R^2}l^2{\Bigg)}$.
Here the line integral appearing in expression (4) has to be calculated along $C$ and $\tilde{C}$, because
we have two different trajectories.

Clearly, the interference term at the detection point, $Q$, is

{\setlength\arraycolsep{2pt}\begin{eqnarray}
I = \cos\left\{{1\over\hbar}\int_{(C)}\sqrt{2m[E - V(\vec{r})]}d\tilde{L} -
{1\over\hbar}\int_{(\tilde{C})}\sqrt{2m[E - V(\vec{r})]}d\tilde{L}\right\}.
\end{eqnarray}

Let us now consider the following trajectories. $C$ is defined as follows, the
particle begins at point $P$, then moves horizontally to point $A$, and finally, 
in vertical form to point $Q$, which is the detection point. On the other hand,
$\tilde{C}$ comprises the fo\-llow\-ing cases, it also starts at $P$, but it moves, vertically, to point
$B$, and afterwards, horizontally to $Q$. We also assume that the $l$--coordinate
of point $P$ is zero, i.e., $l_P = 0$, the horizontal distance between points $B$ and
$Q$, and between points $P$ and $A$, is denoted by $L$, and finally $l_Q$ is the $l$--coordinate of $Q$.
\bigskip

Under these conditions we obtain

{\setlength\arraycolsep{2pt}\begin{eqnarray}
I = \cos{\Bigg[}{L\over\hbar}{\Big(}2mE + 2G_{\infty}{m^2M\over R}\left[1 + \alpha + {\alpha R\over\lambda}({R\over 2\lambda} - 1)\right] \nonumber\\
+ 2G_{\infty}{m^2M\over R}\left[{\alpha R\over 2\lambda^2} - {1 + \alpha\over R}\right]l_Q
+ 2G_{\infty}{m^2M\over R^3}\left[1 + \alpha\right]l_Q^2{\Big)}^{1\over2} \nonumber\\
- {L\over\hbar}\left(2mE + 2G_{\infty}{m^2M\over R}\left[1 + \alpha + {\alpha R\over\lambda}({R\over 2\lambda} - 1)\right]\right)^{1\over2}{\Bigg]}.
\end{eqnarray}

This last expression can be rewritten as 

{\setlength\arraycolsep{2pt}\begin{eqnarray}
I =\cos\left\{-{gm^2Ll_Q\Lambda\over\hbar^2}\left[1 - {\alpha R^2\over 2\lambda^2(1 + \alpha)} - {l_Q\over R}\right]\right\}.
\end{eqnarray}

Here $\Lambda$ denotes the initial reduced wavelength of the particles,
we have also used the fact that $g_{\infty} = g/(1 + \alpha)$, where $g = {GM\over R^2}$.

Imposing the condition $\alpha = 0$ enables us to rewrite expression (7) as

{\setlength\arraycolsep{2pt}\begin{eqnarray}
I_N =\cos\left\{-{gm^2Ll_Q\Lambda\over\hbar^2}\left[1 -{l_Q\over R}\right]\right\},
\end{eqnarray}

\noindent which is the interference term that appears in the case of Newtonian gravity [25].
As a matter of fact, the result of COW does not contain the term that is quadratic in $l_Q$,
in our result it appears because we have introduced a less restricted approximation,
to derive the results of COW we need only a homogeneous Newtonian gravitational field [25], and expression (3) includes
the case of an inhomogeneous gravitational field, i.e., the term
${1 + \alpha\over R^2}l^2$. In other words, expression (8) is the interference term
when we consider, in a Newtonian field, dependence, in the height above the surface of
the Earth, up to second order in $l$.

We may now calculate the difference between the Newtonian and non--Newtonian cases,
and divide this result by the Newtonian value, the outcome reads (approximately)

{\setlength\arraycolsep{2pt}\begin{eqnarray}
\Delta = {\alpha R^2\over 2\lambda^2(1 + \alpha)}\left(1 + {l_Q\over R}\right).
\end{eqnarray}}

Introducing the expression for the Compton wavelength of our new quantum particle
with mass $m_5$ ($\lambda = {\hbar\over cm_5}$), we find that

{\setlength\arraycolsep{2pt}\begin{eqnarray}
\Delta = {\alpha (Rcm_5)^2\over 2\hbar^2(1 + \alpha)}\left(1 + {l_Q\over R}\right).
\end{eqnarray}

We may rewrite (7) as follows

{\setlength\arraycolsep{2pt}\begin{eqnarray}
I = I_N\cos\left\{{gm^2Ll_Q\Lambda\alpha R^2\over 2\hbar^2\lambda^2(1 + \alpha)}\right\}
\pm\sqrt{1 - I_N^2}\sin\left\{{gm^2Ll_Q\Lambda\alpha R^2\over 2\hbar^2\lambda^2(1 + \alpha)}\right\}.
\end{eqnarray}
\bigskip

\section{Quantum Measurements}
\bigskip

As has already been mentioned, in the attempts to solve the quantum measurement problem we may find RPIF [23].
This formalism explains a continuous quantum measurement with the introduction of a restriction on
the integration domain of the corresponding path integral.
This last condition can also be reformulated in terms
of a weight functional that has to be considered in the path integral.

Let us explain this point a little bit better, and suppose that we have a particle which shows one--dimensional movement.
The amplitude $A(q'', q')$ for this particle to move from the point $q'$ to the point $q''$ is called propagator.
It is given by Feynman [26]

\begin{equation}
A(q'', q') = \int d\left[q\right]\exp\left({i\over \hbar}S[q]\right),
\end{equation}

\noindent here we must integrate over all the possible trajectories $q(t)$, $S[q]$ is the action of the system, which is
defined as

\begin{equation}
S[q] = \int_{t'}^{t''}dtL(q, \dot{q}).
\end{equation}

Let us now suppose that we continuously measure the position of this particle,
such that we obtain as measurement ouput a certain function $a(t)$. In other words, the measuring process gives the value $a(t)$
for the \-coor\-di\-na\-te $q(t)$ at each time $t$, and this output has associated a certain error $\Delta a$, which is determined by the
experimental resolution of the measuring device. The amplitude $A_{[a]}(q'', q')$ can be now thought of as a probability amplitude for the continuous measuring process to give the result $a(t)$.
Taking the square modulus of this amplitude allows us to find the probability density for different measurement outputs.

Clearly, the integration domain in the Feynman path--integral should be restricted to those trajectories that match with the experimental output.
RPIF says that this condition can be introduced by means of a weight functional $\omega_a[q]$ [23].
This means that expression (12) becomes now

\begin{equation}
A_a = \int d\left[q\right]\omega_{a}[q]\exp\left({i\over \hbar}S[q]\right).
\end{equation}

The more probable the trajectory $[q]$ is, according to the output $a$, the bigger that $\omega_a[q]$ becomes [23]. This means that the value of $\omega_a[q]$ is approximately one for all trajectories $[q]$ that agree with the measurement
output $a$, and it is almost 0 for those that do not match with the result of the experiment. Clearly, the weight functional contains
all the information about the interaction between measuring device and measured system.

Let us now consider the propagator of a particle whose Lagrangian is given by (3),
(the particle goes from point $P$ to point $Q$)

{\setlength\arraycolsep{2pt}\begin{eqnarray}
U(Q,\tau'';P, \tau') = \left({m\over 2\pi i\hbar T}\right)
\exp\left\{{im\over 2\hbar T}L^2\right\}\int d[l(t)]\exp{\Bigg(}{i\over\hbar}\int_{\tau'}^{\tau''}
{\Bigg[}{m\over 2}\dot{l}^2 \nonumber\\
+~G_{\infty}{mM\over R}
\left(\left[1 + \alpha + {\alpha R\over\lambda}({R\over 2\lambda} - 1)\right]
 - \left[{1 + \alpha\over R} - {\alpha R\over 2\lambda^2}\right]l + {1 + \alpha\over R^2}l^2\right){\Bigg]}dt{\Bigg)}.
\end{eqnarray}}

We now introduce a measuring process, namely we will monitor continuously the
l--coordinate of the particle. Then expression (15) becomes now

{\setlength\arraycolsep{2pt}\begin{eqnarray}
U_{[a(t)]}(Q,\tau'';P, \tau') = \left({m\over 2\pi i\hbar T}\right)
\exp\left\{{im\over 2\hbar T}L^2\right\}\nonumber\\
\times\int d[l(t)]w_{[a(t)]}[l(t)]\exp{\Bigg(}{i\over\hbar}\int_{\tau'}^{\tau''}
{\Bigg[}{m\over 2}\dot{l}^2 \nonumber\\
+~G_{\infty}{mM\over R}\left(\left[1 + \alpha + {\alpha R\over\lambda}
({R\over 2\lambda} - 1)\right] - \left[{1 + \alpha\over R} - {\alpha R\over 2\lambda^2}\right]l + {1 + \alpha\over R^2}l^2\right){\Bigg]}dt{\Bigg)}.
\end{eqnarray}}

The modulus square of this last expression gives the probability of obtaining
as measurement output (for the $l$--coordinate) function $a(t)$.
The weight functional $w_{[a(t)]}[l(t)]$ contains the information concerning the measurement, and
is determined by the experimental construction [23].

At this point, in order to obtain theoretical predictions, we must choose a
par\-ti\-cu\-lar expression for $w_{[a(t)]}[l(t)]$.
We know that the results coming from a Heaveside weight functional [27] and those
coming from a gaussian one [28] coincide up to the order of magnitude.
These last remarks allow us to consider a gaussian weight functional as an approximation of the correct expression.

It will be supposed that the weight functional of our measuring device has
precisely this gaussian form. We may wonder if this is not an unphysical assumption,
and in favor of this argument we may comment that recently it has been proved that
there are measuring apparatuses which show this kind of behaviour [29].

Therefore we may now choose as our weight functional the following expression

\begin{equation}
\omega_{[a(t)]}[l(t)] = \exp\left\{-{2\over T\Delta a^2}\int _{\tau '}^{\tau ''}[l(t) - a(t)]^2dt\right\},
\end{equation}
\bigskip

\noindent here $\Delta a$ represents the error in our measurement.

Hence with the introduction of a continuous quantum measurement the new pro\-pa\-ga\-tor is

{\setlength\arraycolsep{2pt}\begin{eqnarray}U_{[a(t)]}(Q,\tau'';P, \tau') = \left({m\over 2\pi i\hbar T}\right)
\exp\left\{{im\over 2\hbar T}L^2\right\}\nonumber\\
\int d[l(t)]\exp\left\{-{2\over T\Delta a^2}\int_{\tau '}^{\tau ''}[l(t) - a(t)]^2dt\right\}
\exp{\Bigg(}{i\over\hbar}\int_{\tau'}^{\tau''}
{\Bigg[}{m\over 2}\dot{l}^2 \nonumber\\
+ G_{\infty}{mM\over R}\left\{\left[1 + \alpha + {\alpha R\over\lambda}({R\over 2\lambda} - 1)\right] 
- \left[{1 + \alpha\over R} - {\alpha R\over 2\lambda^2}\right]l + {1 + \alpha\over R^2}l^2\right\}{\Bigg]}dt{\Bigg)}.
\end{eqnarray}}

It can be rewritten as follows

{\setlength\arraycolsep{2pt}\begin{eqnarray}
U_{[a(t)]}(Q,\tau'';P,\tau') = \left({m\over 2\pi i\hbar T}\right)
\exp\left\{{im\over 2\hbar T}L^2\right\}\nonumber\\
\times\exp\left\{{i\over\hbar}g_{\infty}mR
\left[1 + \alpha + {\alpha R\over\lambda}({R\over 2\lambda} - 1)\right]T\right\}
\exp\left\{-{2\over T\Delta a^2}\int_{\tau'}^{\tau''}a^2(t)dt\right\}\nonumber\\
\times\int d[l(t)]\exp\left\{{i\over\hbar}\int_{\tau'}^{\tau''}\left[{m\over 2}\dot{l}^2 + 
F(t)l -{m\over 2}\omega^2l^2\right]dt\right\}.
\end{eqnarray}}

In this last expression we have introduced the following definitions, namely 
$F(t) = mg[{\alpha R^2\over 2\lambda^2(1 + \alpha)} - 1]
- {4i\hbar\over T\Delta a^2}a(t)$, and $\omega = i\Omega$, where $\Omega^2 = 2{g\over R}(1 + {2i\hbar R\over mgT\Delta a^2})$.
It is rea\-di\-ly seen that we have now the propagator of a driven harmonic oscillator, but now
frequency and driving force have nonvanishing imaginary parts, which emerge as a direct consequence of our measuring process. We already know how to evaluate this kind
of path integrals [30].
\bigskip

The result of the integration yields (here we do not assume that $l_P = 0$)

{\setlength\arraycolsep{2pt}\begin{eqnarray}
U_{[a(t)]}(Q,\tau'';P, \tau') = \left({m\over 2\pi i\hbar T}\right)^{{3\over2}}
\sqrt{{\Omega T\over\sinh(\Omega T)}}
\exp\left\{{im\over 2\hbar T}L^2\right\}\nonumber\\
\times\exp\left\{{i\over\hbar}g_{\infty}mR
[1 + \alpha + {\alpha R\over\lambda}({R\over 2\lambda} - 1)]T\right\}
\exp\left\{-{2\over T\Delta a^2}\int_{\tau'}^{\tau''}a^2(t)dt\right\}\nonumber\\
\times\exp{\Big(}{im\Omega\over 2\hbar\sinh(\Omega T)}
{\Big[}(l^2_Q + l^2_P)\cosh(\Omega T) - 2l_Ql_P \nonumber\\
-{8i\hbar\over mT\Omega\Delta a^2}\left\{l_QF^{(1)}(\tau'', \tau') + l_PF^{(2)}(\tau'', \tau')\right\} \nonumber\\
+ 2g{l_Q + l_P\over\Omega^2}\left\{{\alpha R^2\over 2\lambda^2(1 + \alpha)} - 1\right\}\left\{\cosh(\Omega T) - 1\right\} \nonumber\\
- 2{g^2\over\Omega^2}\left\{{\alpha R^2\over 2\lambda^2(1 + \alpha)} - 1\right\}^2\left\{{1- \cosh(\Omega T)\over \Omega^2} + {T\sinh(\Omega T)\over 2\Omega}\right\} \nonumber\\
+ {8i\hbar g\over m\Omega^2 T\Delta a^2}\left\{{\alpha R^2\over 2\lambda^2(1 + \alpha)} - 1\right\}\int_{\tau'}^{\tau''}
F^{(1)}(\tau, \tau')\sinh(\Omega(\tau'' - \tau))d\tau\nonumber\\
+ {8i\hbar g\over mT\Delta a^2\Omega^3}\left\{{\alpha R^2\over 2\lambda^2(1 + \alpha)} - 1\right\}\left\{F^{(3)}(\tau'', \tau') - F^{(2)}(\tau'', \tau')\right\}\nonumber\\
+ {32\hbar^2\over m^2T^2\Omega^2\Delta a^4}F^{(4)}(\tau'', \tau'){\Big]}{\Big)},
\end{eqnarray}}

\noindent where $F^{(1)}(\tau'', \tau') = \int_{\tau'}^{\tau''}a(\tau)\sinh(\Omega(\tau - \tau'))d\tau$, we also have defined,
$F^{(2)}(\tau'', \tau') = \int_{\tau'}^{\tau''}a(\tau)\sinh(\Omega(\tau'' - \tau))d\tau$, $F^{(3)}(\tau'', \tau') = 
\int_{\tau'}^{\tau''}a(\tau)\sinh(\Omega(\tau'' - \tau))\cosh(\Omega(\tau - \tau'))d\tau$, and finally
$F^{(4)}(\tau'', \tau') = \int_{\tau'}^{\tau''}d\tau\int_{\tau'}^{\tau}dsa(\tau)a(s)\sinh(\Omega(\tau'' - \tau))\sinh(\Omega(s - \tau'))$. 
\bigskip

\section{Discussion}
\bigskip

Expression (11) shows clearly that the interference pattern emerging in the case of a 
non--Newtonian gravity theory, here the deviation comprises a Yukawa term, does not match
with the results of the inverse--square law situation. If we consider the experimental parameters
of COW, and also the values $\alpha \sim 10^{-3}$ and $\lambda \sim 10^4$cm [31] (here $\lambda$ denotes the range of the Yukawa interaction),
then we deduce that ${gm^2Ll_Q\Lambda\alpha R^2\over 2\hbar^2\lambda^2(1 + \alpha)}\sim 10^{7}(cm)^{-1}l_Q$.
Hence (11) reduces to

{\setlength\arraycolsep{2pt}\begin{eqnarray}
I = I_N\cos\left\{10^{7}(cm)^{-1}l_Q\right\}
\pm\sqrt{1 - I_N^2}\sin\left\{{10^{7}(cm)^{-1}l_Q}\right\}.
\end{eqnarray}

The dependence in $l_Q$ of (21) could, in principle, be  detected.

{\setlength\arraycolsep{2pt}\begin{eqnarray}
{\Delta I\over\Delta l_Q} \sim 10^{7}(cm)^{-1}\left(-I_N\sin\left\{10^{7}(cm)^{-1}l_Q\right\}
\pm\sqrt{1 - I_N^2}\cos\left\{{10^{7}(cm)^{-1}l_Q}\right\}\right)\nonumber\\
+ {\Delta I_n\over \Delta l_Q}\left(\cos\left\{10^{7}(cm)^{-1}l_Q\right\} \mp{I_N\over\sqrt{1 - I_N^2}}\sin\left\{{10^{7}(cm)^{-1}l_Q}\right\}\right).
\end{eqnarray}

Knowing that these Yukawa terms emerge in a natural manner in some unified field theories
[32] (they are related to the existence of an intermediate--range new force coupled to baryon number or hypercharge), 
then our result could help to determine the phenomenological parameters $\alpha$ and $\lambda$.
Parameter $\alpha$ could also be composition--dependent [17], this possibility
could also be tested employing expression (11), i.e., performing the interference experiment with different type
of materials. The relevance of this last point is related to the analysis of the validity of WEP and of the Strong Equivalence Principle (SEP),
namely a composition--independent $\alpha$ would not violate WEP, but it might violate SEP [33].

Let us now consider the case in which our two beams start at point $P$ and afterwards
are detected at point $Q$, here we do not consider any restriction on the wavelength of
the beams. We also assume that the $l$--coordinate of the beams is being continuously measured,
in other words, we may use expression (20) to calculate the corresponding wave function.
The measuring process will take place under the following restrictions: (i) two functions
are obtained as measurement outputs, namely $a(t)$ and $b(t)$ (each beam has its own function);
(ii) we carry out this experiment using two devices (each beam has its own measuring device), whose errors are not the same, i.e.,
$\Delta a \not = \Delta b$. We may then calculate the emerging interference pattern, the result is given
by the real part of the following expression

{\setlength\arraycolsep{2pt}\begin{eqnarray}
I = \exp{\Bigg(}{im\over 2\hbar}
{\Bigg[}(l^2_Q + l^2_P)\left\{{\Omega\over\tanh(\Omega T)} - {\Gamma\over\tanh(\Gamma T)}\right\}\nonumber\\
+ 2l_Ql_P\left\{{\Gamma\over\sinh(\Gamma T)} - {\Omega\over\sinh(\Omega T)}\right\}
+ {8i\hbar\over mT}
{\Big(}{l_QF^{(1)}(\tau'', \tau') + l_PF^{(2)}(\tau'', \tau')\over\sinh(\Omega T)\Delta a^2}\nonumber\\
- {l_Qf^{(1)}(\tau'', \tau') + l_Pf^{(2)}(\tau'', \tau')\over\sinh(\Gamma T)\Delta b^2}{\Big)} \nonumber\\
+ 2g\tilde{\alpha}
{\Bigg[}\left\{{\cosh(\Omega T) -1\over\Omega\sinh(\Omega T)}\right\}
{\Big(}l_Q + l_P
+ {g\over\Omega^2}\tilde{\alpha}{\Big)} - {T\over 2\Omega^2}\nonumber\\
-\left\{{\cosh(\Gamma T) -1\over\Gamma\sinh(\Gamma T)}\right\}
{\Big(}l_Q + l_P 
+ {g\over\Gamma^2}\tilde{\alpha}{\Big)} + {T\over 2\Gamma^2}{\Bigg]}\nonumber\\
+ {8i\hbar g\over mT}\tilde{\alpha}
{\Big[}{F^{(3)}(\tau'', \tau') - F^{(2)}(\tau'', \tau')\over\Delta a^2\sinh(\Omega T)\Omega^2}
- {f^{(3)}(\tau'', \tau') - f^{(2)}(\tau'', \tau')\over\Delta b^2\sinh(\Gamma T)\Omega^2}\nonumber\\
+ {F^{(5)}\over \Omega\Delta a^2 sinh(\Omega T)} - {f^{(5)}\over \Gamma\Delta b^2 sinh(\Gamma T)}{\Big]}\nonumber\\
+ {32\hbar^2\over m^2T^2}\left\{{F^{(4)}(\tau'', \tau')\over\Omega\sinh(\Omega T)\Delta a^4}
- {f^{(4)}(\tau'', \tau')\over\Gamma\sinh(\Gamma T)\Delta b^4}\right\}{\Bigg]}{\Bigg)}.
\end{eqnarray}}

Here $\Gamma^2 = 2{g\over R}(1 - {2i\hbar R\over mgT\Delta b^2})$.
Additionally, the following functions have been employed $f^{(1)}(\tau'', \tau') = \int_{\tau'}^{\tau''}b(\tau)\sinh(\Gamma(\tau - \tau'))d\tau$$f^{(2)}(\tau'', \tau') =
\int_{\tau'}^{\tau''}b(\tau)\sinh(\Gamma(\tau'' - \tau))d\tau$,
$f^{(3)}(\tau'', \tau') = \int_{\tau'}^{\tau''}b(\tau)\sinh(\Gamma(\tau'' - \tau))\cosh(\Gamma(\tau - \tau'))d\tau$.
Also the following parameters have been defined $f^{(4)}(\tau'', \tau') = \int_{\tau'}^{\tau''}d\tau\int_{\tau'}^{\tau}dsb(\tau)b(s)\sinh(\Gamma(\tau'' - \tau))\sinh(\Gamma(s - \tau'))$,
$F^{(5)} = \int_{\tau'}^{\tau''}F^{(1)}(\tau, \tau')\sinh(\Omega(\tau'' - \tau))d\tau$,
$f^{(5)} = \int_{\tau'}^{\tau''}f^{(1)}(\tau, \tau')\sinh(\Gamma(\tau'' - \tau))d\tau$,
and finally $\tilde{\alpha} = {\alpha R^2\over 2\lambda^2(1 + \alpha)} - 1 $.
\bigskip

It is readily seen that the emerging interference pattern depends not only on the parameters
that define the Yukawa interaction, $\alpha$, $\lambda$, but also on the parameters that appear in RPIF,
namely $a(t)$, $b(t)$, $\Delta a$, and also on $\Delta b$. Hence we could compare with the Newtonian situation [24],
and therefore obtain an additional scheme that could render some restrictions upon the possible values of $\alpha$ and $\lambda$.
If we consider the limit $\alpha \rightarrow 0$ and $\lambda \rightarrow \infty$,
we obtain the results of the Newtonian situation [24].

Taking a look at expression (23) we may notice that that the mass of the test particle appears, explicitly, in the
expression for the interference pattern, as happens also in the context of Newtonian gravity [25], and
always as a function of $\hbar/m$.

Expression (23), at the same time, also gives a new testing framework for the theoretical predictions of RPIF,
which makes the work in this direction a little bit more complete [34].

This procedure could also be applied to some other possible modifications of the Newtonian gravity law, for instance, we could consider the case of the extra potential $V(r) = \alpha/r^5$ (this kind of terms arise in
some modified gravitational theories in which the non--Newtonian behavior stems from
antisymmetric terms in the metric tensor [35]), and then carry out the same analysis,
but now using as Lagrangian

{\setlength\arraycolsep{2pt}\begin{eqnarray}
L = {m\over 2}\dot{\vec{r}}^2 + {1\over R}\left(GmM + {\alpha \over R^4}\right)
- \left({GmM\over R} + 5{\alpha \over R^5}\right){l\over R} + \left({GmM\over R} + 15{\alpha \over R^5}\right){l^2\over R^2}.
\end{eqnarray}

At this point the feasibility of the present proposal must be addressed.
A monocromatic beam of particles could be used (as in COW case [22]), but here no
restriction on the size of the corresponding wave packets is needed, in COW
this condition emerges because a WKB approach describes the physical situation.
Afterwards, the beam could be split in two parts, and then the vertical coordinate
(in the present work it has been denoted by $l$) of each one of these beams has
to be continuously monitored, and finally, both beams have to be brought together,
and the corresponding inter\-fe\-rence pattern measured. In contrast to most of the
existing experiments [9, 14, 17], the present idea could allow us to detect
the effects of a Yukawa term upon quantum systems.
Nevertheless, experimentally, the continuous monitoring of a moving particle lies, currently, outside the present technological
posibilities [36]. Ne\-ver\-withstanding, the advances that in the topic of trapped ions have been achieved [37], allows us to consider the possibility of carrying out the needed experiments in a, hopefully, near future.

As was mentioned before, the ``geophysical window'' has not been tested severely [13], i.e., over the
10m--1000m distance scale. The lower limit of this range could be used in the present
kind of proposals. Of course, regions smaller than 10m could also be explored, for instance,
the 1cm--100cm distance scale. If we consider $\lambda \sim 10$m, then the current experimental limit reads
$\alpha \sim 10^{-1}$ [17]. Looking at expression (23) we may notice that a crucial point
in this kind of proposals concerns the resolution of the measuring devices, i.e.,
$\Delta a$ and $\Delta b$. Hence one of the points that determines the feasibility
of the present idea is related to a, enoughly, small experimental error. The situation
in which $\Delta a \sim 2 \mu$m (which is the resolution in the case of a particle
in a Paul trap [38], and therefore at least in the context of motionless
situations has already been achieved) and $\Delta b = 10^3\times \Delta a$ could be an interesting case to consider.

Finally, a word must be said about some schemes which also imply a violation
of the equivalence principle [39]. In these models the contradictions (between the predictions of general relativity and the results in a Minkowskian
spacetime) emerge as a purely quantum mechanical effect, i.e., they appear when
a quantum system, the one has no classical counterpart, is embeded in a curved spacetime [40, 41]. This should be no surprise, indeed, even 
the kinematical description of quantum mechanical systems (without classical analogue) moving in a
classical gravitational field shows conceptual difficulties [42], for instance,
there is no consistent definition for the concept of time of flight probability distribution.
This simple example also shows, very clearly, the danger of extrapolating,
to quantum systems without classical analogue, the concepts of classical physics.
Of course, these last arguments do not imply that the analysis of the quantum
effects in a classical gravitational field shall not be carried out, they only assert
(as it has already been pointed out [43]) that one should be very careful when addressing this issue.

In the present work, the possible violations of the equivalence principle are due to the presence of an additional interaction (the Yukawa term), a factor that is absent in the
aforementioned models [39, 40, 41].

An interesting question at this point is the following one: let us suppose that we have performed
the here proposed experiment, and that it implies a violation of the equivalence
principle, how could we determine if this violation stems from the existence
of a Yukawa term or it is quantum induced?, here the phrase ``quantum induced'' means the effects of
a classical gravitational field upon a quantum mecha\-ni\-cal system without classical analogue.
The answer might come from the role that mass plays in the measurement outputs.
Indeed, if we take a look at the dependence upon mass of the interference pattern, expression (23), it can be readily seen that it is
not the same dependence that appears in some situations in which this violation
is quantum induced (see expression (16) of [40] and also of [41] and hence it could be possible to determine the origin of this
violation, i.e., it would suffice to perform the experiment several times, using each time
a different mass.

Though the COW experiment has been performed with a very good precision [22, 44], currently 
there are some works which endow COW with some discrepancies [45]. Concerning these
new effects, possibly the mixture of a quantum measurement process and gravitational effects renders an unavoidable modification to the de
Broglie's wave--particle duality [46]. Of course, more work is needed in this direction,
where not only the case of quantum demolition measurements (as position monitoring) have to been considered, but also the possibilities that quantum
nondemolition measurements [47] could offer in this issue have to be analyzed.

\bigskip

\Large{\bf Acknowledgments}\normalsize
\bigskip

The author would like to thank A. A. Cuevas--Sosa and A. Camacho--Galv\'an for their help,
and D.-E. Liebscher for the fruitful discussions on the subject.
The hospitality of the Astrophysikalisches Institut Potsdam is also kindly acknowledged.
It is also a pleasure to thank R. Onofrio for bringing Ref. [15] to my attention. The author is also indebeted
to D. V. Ahluwalia for providing a copy of Ref. [44] before its publication, and for the
personal communication of reference [43].
This work was supported by CONACYT (M\'exico) Posdoctoral Grant No. 983023.

\end{document}